\documentclass[preprint,preprintnumbers,amsmath,amssymb,superscriptaddress]{revtex4}

\usepackage{graphicx}
\usepackage{dcolumn}
\usepackage{bm}

\begin{document}

\preprint{APS/123-QED}

\title{Electrons on the surface of Bi$_2$Se$_3$ form a topologically-ordered two dimensional gas with a non-trivial Berry's phase}

\author{Y. Xia}
\affiliation{Joseph Henry Laboratories of Physics, Department of Physics, Princeton
University, Princeton, NJ 08544, USA}
\author{L. Wray}
\affiliation{Joseph Henry Laboratories of Physics,  Department of Physics, Princeton
University, Princeton, NJ 08544, USA}
\author{D. Qian}
\affiliation{Joseph Henry Laboratories of Physics,  Department of Physics, Princeton
University, Princeton, NJ 08544, USA}
\author{D. Hsieh} \affiliation{Joseph Henry Laboratories of Physics,  Department of Physics, Princeton
University, Princeton, NJ 08544, USA}
\author{A. Pal}
\affiliation{Joseph Henry Laboratories of Physics, Department of Physics, Princeton
University, Princeton, NJ 08544, USA}
\author{H. Lin}
\affiliation{Department of Physics, Northeastern University, Boston, MA}
\author{A. Bansil}
\affiliation{Department of Physics, Northeastern University, Boston, MA}
\author{D. Grauer}
\affiliation{Department of Chemistry, Princeton University, Princeton, NJ 08544, USA}
\author{Y. S. Hor}
\affiliation{Department of Chemistry, Princeton University, Princeton, NJ 08544, USA}
\author{R. J. Cava}
\affiliation{Department of Chemistry, Princeton University, Princeton, NJ 08544, USA}
\author{M. Z. Hasan}
\affiliation{Joseph Henry Laboratories of Physics,  Department of Physics, Princeton
University, Princeton, NJ 08544, USA}
\affiliation{Princeton Center for Complex Materials, Princeton
University, Princeton, NJ 08544, USA}
  \email{mzhasan@Princeton.edu}

\date{\today}



\maketitle

\textbf{Recent experiments and theories suggest that spin-orbit coupling (SOC) can lead to new phases of quantum matter with highly non-trivial collective quantum effects. Two such phases are the quantum spin Hall insulator \cite{kane05, konig07} and the strong topological insulator (STI) \cite{hsieh08, fu07, zhang} both realized in the vicinity of a Dirac point \cite{hsieh08} but yet quite distinct from graphene \cite{graphene}. The STI phase contains unusual two-dimensional edge states, which may provide a route to spin-charge separation in higher dimensions \cite{qi08, ran08} and may realize other novel magnetic and electronic properties \cite{moore08}. It has been suggested that the interface of an STI and a conventional superconductor can support exotic quasiparticle states which could be used in fault-tolerant computing schemes \cite{fu2008}. It is currently believed that the Bi$_{1-x}$Sb$_{x}$ alloys realize the only known topological insulator phase in the vicinity of a 3D Dirac point\cite{hsieh08}. However, a particular challenge for the topological insulator Bi$_{1-x}$Sb$_{x}$ system is that the insulating bulk gap is small and the material contains a significant amount of alloying disorder which is difficult to gate for the manipulation and control of charge carriers to realize a device. The topological insulator Bi$_{1-x}$Sb$_{x}$ features five surface states of which only one carries the topological quantum number. Therefore, there is an extensive world-wide search for similar topological phases in stoichiometric materials with no alloying disorder, a larger gap and fewer yet odd numbered surface states. Here we present angle-resolved photoemission (ARPES) data and electronic structure calculation which seems to suggest the existence of a Z$_2$ topological phase with a surface Berry's phase in the stoichiometric compound Bi$_2$Se$_3$. Our results taken collectively suggest that the undoped Bi$_2$Se$_3$ can serve as the parent matrix compound for the long-sought topological materials with only one surface state where in-plane carrier transport would be fully dominated by the Z$_2$ topological effects. Moreover, the undoped compound should exhibit topological quantum effects at room temperature.}


Undoped Bi$_2$Se$_3$ is a semiconductor which belongs to the class of materials Bi$_2$X$_3$ (X=S,Se,Te) with a rhombohedral crystal structure (space group D$^5_{3d}$ ($R\bar{3}m$)). The unit cell contains five atoms, with the quintuple layers ordered in the Se(1)-Bi-Se(2)-Bi-Se(1) fashion. These compounds have been studied extensively in connection to thermoelectric applications \cite{disalvo99}. Experimental measurements suggest a band gap of approximately 0.35 eV \cite{black57, wyckoff86, shroeder73} where as the bulk band calculations estimate the gap to be in between 0.24 to 0.32 eV \cite{larson02, mishra96}. Various measurements and calculations have reported that although the bulk of the material is a direct gap semiconductor \cite{greanya02} its electronic properties can vary significantly with different sample preparation conditions \cite{hyde74}, with a strong tendency to be n-type \cite{greanya02} due to impurity and vacancies.

Single crystal of Bi$_2$Se$_3$ was grown by melting stoichiometric
mixtures of high purity elemental Bi and Se in a 4 mm inner
diameter quartz tube. The sample was cooled over a period of two
days, from 850 to 650 $^{\circ}$C, and then annealed at that
temperature for a week. Single crystals were obtained and could be easily cleaved from the boule. High-resolution ARPES measurements were then performed using 17 to 45eV photons on beamline 5-4 at the Stanford Synchrotron Radiation Laboratory (SSRL) and beamline 12.0.1 of the Advanced Light Source
at the Lawrence Berkeley National Laboratory. The energy and momentum resolutions were 15meV and 2\% of the surface Brilloin Zone (BZ) respectively using a Scienta analyzer. The samples were cleaved \emph{in situ} between 10K and 55K under pressures of less than $5\times 10^{-11}$ torr, resulting in shiny flat surfaces. For the deposition-effect measurements, the Fe atoms were deposited using an e-beam heated evaporator at a rate of approximately 0.12$\AA$/min. We also carried our $\textit{surface}$ band calculations for the Bi$_2$Se$_3$(111) surface for comparison with the experimental data. The calculations were performed with the LAPW method in slab geometry using the WIEN2K package \cite{blaha01}. GGA of Perdew, Burke, and Ernzerhof
\cite{perdew96} was used to describe the exchange-correlation potential. SOC was included as a second variational step using scalar-relativistic eigenfunctions as basis after the initial calculation is converged to self-consistency. The surface was simulated by placing a slab of twelve quintuple layers in vacuum. A grid of $21\times 21 \times 1$ points were used in the calculations, equivalent to 48 k-points in the irreducible BZ and
300 k-points in the first BZ.

\begin{figure}
\includegraphics[width=0.5\textwidth]{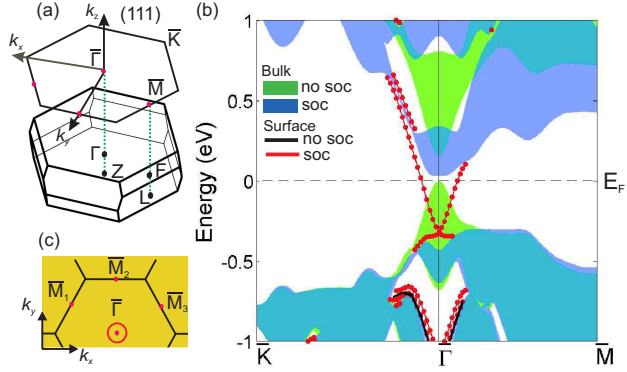}
\caption{\label{fig:theory} Spin-orbit interaction induced surface Fermi
surface: (a) A schematic of the bulk 3D BZ of Bi$_2$Se$_3$ and the 2D
BZ of the projected (111) surface. (b) LDA band-structure of the 2D surface states along the $\bar{K}-\bar{\Gamma}-\bar{M}$ k-space cut. Bulk band projections are represented by the shaded areas. The band-structure results with spin-orbit coupling (SOC) is presented in blue and that without SOC is in green. No pure surface band is observed within the insulating gap without SOC (black lines). One pure gapless surface band crossing E$_F$ is observed when SOC is included (red lines). (c) The corresponding surface FS is a single ring centered at $\bar{\Gamma}$. The band responsible for this ring is singly degenerate. The time-reversal-invariant momenta (TRIM) on the (111) surface BZ : the $\bar{\Gamma}$ and the three $\bar{M}$ points are marked by red dots.}
\end{figure}

\begin{figure}

\includegraphics[width=0.5\textwidth]{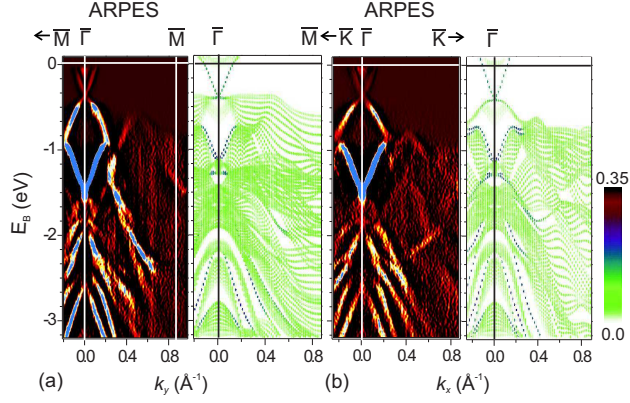}
\caption{\label{fig:cuts} Experimental band dispersions are compared with theoretical calculations: The band second-derivative images and first-principles calculation results along (a) $\bar{\Gamma}-\bar{M}$ and (b) $\bar{\Gamma}-\bar{K}$ directions are presented. The color of the calculated bands represents the fraction of electronic charge residing in the surface layers. A rigid shift of E$_F$ is included to match the lowest energy excitations in the ARPES data with calculations, a consequence of the system being somewhat electron-doped. The strongest signals are observed from the surface states (see text).}
\end{figure}

\begin{figure*}
\includegraphics[width=1.0\textwidth]{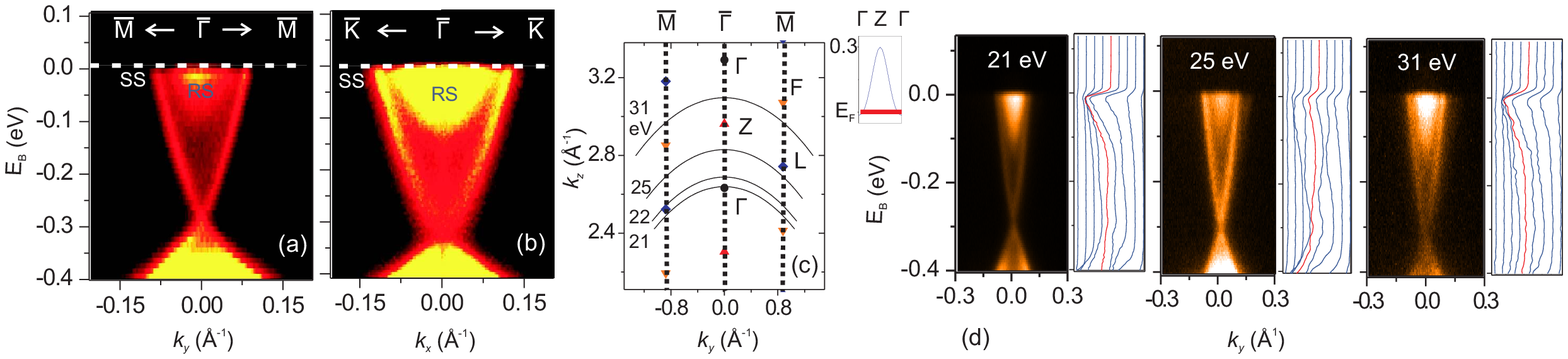}
\caption{\label{fig:energydep} $k_z$ dependence of low-lying
excitations near $\bar{\Gamma}$: High resolution surface
band dispersion data near the $\bar{\Gamma}$ point along the (a)
$\bar{\Gamma}$-$\bar{M}$ and (b) $\bar{\Gamma}$-$\bar{K}$
directions measured at 22eV photon energy are presented. The "V" shaped pure surface state (SS) band is nearly isotropic along both axes, forming a cone in the $k_x-k_y$ space. (c) Results from the energy dependence study is presented along $\bar{\Gamma}$-$\bar{M}$ going from $\Gamma$ at 21 eV to 0.41
$\Gamma-Z$ at 31 eV in 1eV intervals. Solid arcs indicate points in the bulk 3D BZ seen by the detector over a $\theta$ range of $\pm 30^{\circ}$. (d) The energy dispersion data at three selected energies show no $k_z$ dependence for the outer SS band as well as the inner resonant state (RS) feature. At each energy the corresponding energy distribution curve is presented (top). The
calculated bandwidth of the lowest 3D bulk conduction band (inset) is about 0.3 eV which is significantly larger than the instrumental resolution.}
\end{figure*}

Figure ~\ref{fig:theory}(b) shows the calculated surface electronic
structure of the Bi$_2$Se$_3$(111) along the $\bar{K}-\bar{\Gamma}-\bar{M}$ direction. The result with and without SOC are overlaid together for comparison. The bulk band projections are represented by the shaded areas in blue with SOC and green without. In the bulk, time-reversal symmetry demands
$E(\vec{k},\uparrow)=E(-\vec{k},\downarrow)$ while space inversion
symmetry requires that $E(\vec{k},\uparrow)=E(-\vec{k},\uparrow)$.
Therefore, all the energy bands should be doubly-degenerate.
However, space inversion symmetry is broken at the surface, so the
degeneracy of the SS can be broken by spin-orbit interactions. Nevertheless, Kramer's Theorem dictates that spin degeneracy should be preserved at some high symmetry points of the surface BZ that are time reversal invariant momenta (TRIM). In the Bi$_2$Se$_3$(111) surface BZ, these are given by $\bar{\Gamma}$ and three $\bar{M}$ points, located 60$^{\circ}$ away from each other and in the middle of two $\bar{\Gamma}$ points (Fig.~\ref{fig:theory}(c)). The calculated result without SOC shows a
clear gap between the valence and conduction bulk bands, with no pure surface bands crossing E$_F$. SOC drastically changes the band structure of the SS. One finds two singly degenerate surface bands emerging from the bulk projection which are paired together at the $\bar{\Gamma}$ point (one of the Kramers' points). The top surface band forms an electron pocket at E$_F$, giving a ring-like surface Fermi surface (FS). We emphasize that while doubly degenerate states may appear as two overlapping bands in the calculated band structure for some k-values, the two surface bands each appear only once, and represent one eigenstate at a given momentum and most importantly, the central Fermi surface is singly degenerate.

The calculated valence band dispersion is in good correspondence while making a comparison with the experimental data taken in both the
$\bar{\Gamma}-\bar{M}$ and $\bar{\Gamma}-\bar{K}$ directions (Fig.
~\ref{fig:cuts}). The first and twelfth quintuple blocks are taken as the surfaces of the slab. The fraction of electron charge residing in the atomic spheres of these two surface blocks are presented by the color of the bands. Bright lines indicate bands located inside the bulk sample. These bands should overlap with a bulk energy band at a particular $k_z$ value, after
projected onto the 2D surface BZ. A good agreement is seen by shifting the E$_F$ of the calculated result to 20 meV above the bottom of the lowest conduction band - a consequence of doping the semiconductor matrix with electrons. While this observation suggests that the bulk is slightly electron doped (n-type) as observed in many transport measurements, it is an advantage for the fact that we can have spectroscopic (ARPES) look at the large part of the complete band-structure of the surface states which was not possible in fully insulating BiSb \cite{hsieh08}. A complete spectroscopic view is of significant advantage for we would like to study the connectivity of the surface and bulk states at all energies between the valence and conduction bands if possible to determine the unique and specific class of topological order of the parent materials \cite{fu07}.

In the measured spectra, the strongest quasiparticle signals are typically observed near normal electron emission, at around 1.5-2.0 eV. This band corresponds to a pure surface band lying outside the bulk projection of states. Strong quasiparticle signals near $\bar{\Gamma}$ are of surface origin, based on their direct correspondence with the band calculation. Our energy dependent study of the bands suggest insignificant $k_z$ dependence of these state, confirming the 2D (surface) nature of the bands. Near E$_F$, one finds an additional 150 meV electron pocket-like feature inside the spin-split surface band. This feature is a hybrid resonant state formed by the superposition of the surface and the bulk.



To investigate the resonant state (RS) feature centered at $\bar{\Gamma}$, high momentum resolution data is presented in Fig. ~\ref{fig:energydep}. The extracted band velocity of the outer pure SS in both $\bar{\Gamma}$-$\bar{K}$ and $\bar{\Gamma}$-$\bar{M}$ directions are approximately $5\times
10^5$ m/s, close to the calculated values. The SS crosses E$_F$ at
0.09 $\AA^{-1}$ in the $\bar{\Gamma}$-$\bar{M}$ direction and 0.11
$\AA^{-1}$ in the $\bar{\Gamma}$-$\bar{K}$ direction. The
dependence of the bands at three selected energies and their
respective energy distribution curves (EDCs) are presented in
panel (d). By changing the energy of the incident photon, one can
move to different $k_z$ values in the 3D bulk BZ (Fig.
~\ref{fig:energydep}(c)). The inner potential $V_0$ used is
approximately 10 eV, given by the muffin-tin zero of the
calculation. The value is obtained by averaging the atomic
potential over the interstitial region, using a muffin-tin radius
of 2.5 bohr. Moving the photon energy at normal emission from
21eV, corresponding to $\Gamma$ in the bulk 3D BZ, to 31eV,
corresponding to 0.41 $\Gamma-Z$(or -0.59 $\Gamma-Z$ after mapping
into the first BZ), one finds no $k_z$ dispersion of the bands. A
variation of the quasiparticle intensity is however observed with
changing photon energy, due to matrix element effects. The
calculated $\Gamma-Z$ dispersion of the lowest conduction band is
presented in the inset. The resulting bandwidth is approximately
300 meV. Therefore, if the features above 0.15eV are purely due to
the bulk, one should observe a clear dispersion as $k_z$ moves
away from the $\Gamma$ point. The two closest bulk bands - the
highest bulk valence band and the next conduction band, are at
least 500 meV away with bandwidths of approximately 200 meV, so
they are not to affect the quasiparticle signal near $E_F$. This
finding confirms that the outer "V" band as well as the inner
electron pocket-like feature are due to the surface.

\begin{figure}
\includegraphics[width=0.48\textwidth]{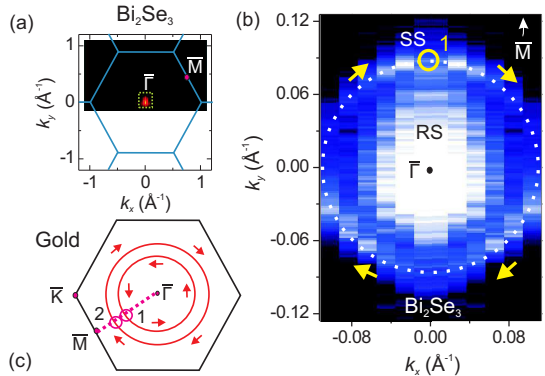}
\caption{\label{fig:fermi} The surface Fermi surface : (a) the surface FS is a small pocket around $\Gamma$. (b) High momentum resolution data around $\bar{\Gamma}$ show a single ring formed by the pure surface state band. In the middle of the ring is a filled-in disk-shaped spectral intensity reflecting resonant states (see text). The pure surface Fermi surface of BiSe is different from that of spin-orbit pair Fermi surface observed on metallic gold Au(111). (c) The Au(111) surface FS, which has two rings (each non-degenerate) surrounding the $\bar{\Gamma}$ TRIM. An electron circling the gold FS can only carry a Berry's phase of 0, characteristic of a trivial topological metal, Z$_2$ ==+1. The single surface Fermi surface observed in Bi$_2$Se$_3$ reflects its non-trivial topological character Z$_2$ =-1.}
\end{figure}

The surface FS is then presented in Figure ~\ref{fig:fermi}. Panel
(a) presents the ARPES data over the entire 2D (111) surface BZ.
No E$_F$ crossings are observed other than the features centered
at $\bar{\Gamma}$. The three TRIMs located at $\bar{M}$ are not
enclosed by any Fermi arcs. This observation is in contrast to the
surface FS of Bi$_{1-x}$Sb$_{x}$ \cite{hsieh08}, which contains a electron-hole pocket features around each of the $\bar{M}$ points. The region
around $\bar{\Gamma}$ is zoomed in with finer momentum resolution
data (panel (b)). One finds a ring-like feature formed by the
outer "V" SS surrounding the filled-in RS disk. This ring is single degenerate thus spin-polarized as expected from the correspondence bewteen the LDA band calculation. An electron encircling the surface FS that encloses a TRIM collects a Berry's phase of $\pi$ mod $2\pi$, picking up a value of $\pi$ from each of the Fermi arcs that enclose the TRIM. At the first sight the surface FS in Bi$_2$Se$_3$ may look like the well-known surface FS in bulk metallic strong spin-orbit coupled gold (Au). However, the SS in gold(111) is split into two parabolic bands each singly degenerate, which are shifted
in momentum-space from each other and both enclose the $\Gamma$-point \cite{hoesch04, lashell96}. The resulting surface FS topology contains two concentric rings around $\bar{\Gamma}$, instead of one as observed in our data for Bi$_2$Se$_3$. Therefore, the Fermi arcs in gold encloses the TRIM an even number of times, with two $E_F$ crossings between a pair $\bar{\Gamma}$ and $\bar{M}$ points. Because each of the rings is singly degenerate (necessarily spin polarized), an electron encircling the FS can only carry a trivial Berry's phase of 0 mod $2\pi$, by picking up one $\pi$ from each of the rings. This makes gold surface states topologically trivial.

\begin{figure}
\includegraphics[width=0.45\textwidth]{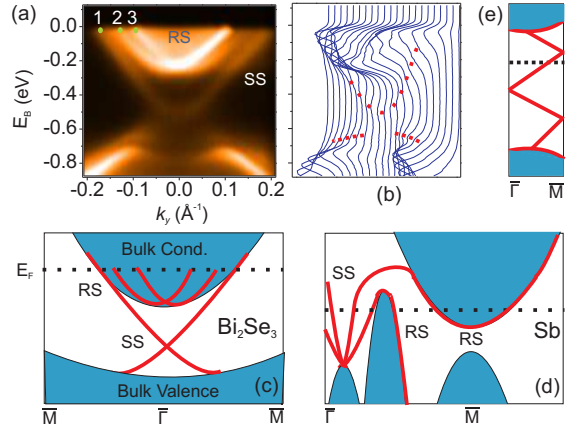}
\caption{\label{fig:distort} The iron deposited surface: band structure in a broken time-reversal setting and the determination of contribution of the RS feature to the Berry's phase counting: (a) a pair of spin-split bands are extended from the RS feature after the chemical potential is shifted with doping. A small gap opens at $\bar{\Gamma}$, due to the loss of time reversal symmetry. The corresponding EDC is shown in (b). Similar to (c) n-Bi$_2$Se$_3$, (d) the band structure of the strong topological metal Sb(111) contains RS at some TRIMs. The pair of spin-split RS bands in Bi$_2$Se$_3$ resembles the SS in Sb. In both cases (e) a "partner-switching" behavior is observed, a critical characteristics in generating an odd number of surface state Fermi crossings.}
\end{figure}

To determine the topological nature of Bi$_2$Se$_3$, one needs to
determine if the RS contributes an odd or even number of crossings
between two TRIMs. In order to determine the nature of the RS states and its connection to the conduction band structure we need to doped the surface of Bi$_2$Se$_3$ with atoms that donates electrons on the surface. This can be done by depositing alkali atoms on the cleaved surface under UHV conditions. For this particular experiment, we have chosen to deposit iron on the surface. Figure ~\ref{fig:distort}(a) presents ARPES data at 29eV along $\bar{\Gamma}-\bar{M}$ after Fe atoms of less than $1\%$ of a monolayer is deposited onto the sample. The system becomes strongly electron doped, with the bottom of the "V" band shifted by more than 200meV. In addition to a shift in the chemical potential, the outer "V" band also becomes less dispersive, with a decreases in the band velocity near $E_F$. More interestingly, the inner RS appears to be the bottom of a pair of spin-split parabolic bands shifted from each other in k-space. These bands are reminiscent of those observed in the SS of Au(111) \cite{hoesch04, lashell96} and Sb(111) \cite{sbarpes} at $\bar{\Gamma}$. The result shows that the RS feature crosses $E_F$ twice, bringing the total number of
crossings between a $\bar{\Gamma}$-$\bar{M}$ pair to three. 


Since iron is magnetic its use in doping the surface also serves a separate but important purpose which has to do with the possiblity of observing the system's response to the breaking of time reversal invariance on the surface.
With magnetic irons on the surface Kramer's degeneracy is broken at the TRIMs and a gap is expected to open up. A small gap or strong spectral weight suppression is indeed observed at $\bar{\Gamma}$ (Fig. ~\ref{fig:distort}(a)). The outer "V" band becomes more parabolic and is detached from the lower "$\wedge$" band. The signal intensity at $\bar{\Gamma}$ becomes very weak, whereas in the undoped case, the signal is strongest at the Dirac point. For this reason, an exact measurement of the gap size is difficult. Spectral suppression or a weak gap opening at the Kramers' point suggest that magnetic field and disorder can cut the k-space surface band thread that connects the bulk valence and conduction bands driving the topological insulator into a trivial band insulator as expected in Z$_2$ theory of these materials
\cite{ryu} 

Therefore, the FS of the Bi$_2$Se$_3$(111) SS only encloses the $\bar{\Gamma}$ point an odd number of times, giving it a $\nu_0=1$ and Z$_2$=-1 band topology similar to Bi$_{1-x}$Sb$_{x}$. However, the surface transport of pure undoped Bi$_2$Se$_3$ would then be dominated by topological effects since there is only one surface band that is also topological. Our calculation suggests that the topological character should be preserved in the undoped compound, which is insulating with metallic boundary states. One might be able to obtain the pure undoped compound by hole-doping the n-type Bi$_2$Se$_3$ thereby shifting the chemical potential downwards. Such a chemical alteration would also remove the resonant state discussed previously, leaving a single ring in the surface FS which carries a $\pi$ Berry's phase. Treated along these lines our data in Fig.-3 and 5 suggest that the undoped material should have a band gap of about 0.3 eV which is sufficient to keep it insulating at room temperature.

In conclusion, we have calculated the band structure of Bi$_2$Se$_3$(111) surface and found that spin-orbit coupling induces a single non-degenerate band crossing at E$_F$. Our experimental data agrees with our surface band calculations as far as the band topology is concerned which demonstrates that the stoichiometric compound belongs to the Z$_2$=-1 topological class. We have also shown that since that Fermi surface contains only one surface state, the transport properties of undoped Bi$_2$Se$_3$ would be dominated by topological effects and since the band gap is large (Fig.-3 and 5) the material can be considered as a room temperature topological insulator without any alloying disorder.


\begin{thebibliography}{99}
\bibitem{kane05} C.L. Kane and E.J. Mele, Phys. Rev. Lett. {\bf 95}, 226801 (2005); B.A. Bernevig and S-C. Zhang, Phys. Rev. Lett. {\bf 96}, 106802 (2006).
\bibitem{konig07} M. Konig \emph{et al.}, Science {\bf 318}, 766 (2007).
\bibitem{hsieh08} D. Hsieh \emph{et al.}, Nature {\bf 452}, 970 (2008).
\bibitem{fu07} L. Fu and C.L. Kane, Phys.\ Rev. B {\bf 76}, 045302 (2007).
\bibitem{zhang} S.-C. Zhang, Physics {\bf 1}, 6 (2008).
\bibitem{graphene} K.S. Novoselov \emph{et al.}, Nature {\bf 438}, 197-200 (2005); Y. Zhang \emph{et al.}, Nature {\bf 438}, 201-204 (2005).
\bibitem{qi08} X.-L. Qi \emph{et al.}, Nature Phys. {\bf 4}, 273 (2008).
\bibitem{ran08} Y. Ran \emph{et al.}, arxiv.org/abs/0801.0627 (2008).
\bibitem{moore08} J.E. Moore \emph{et al.}, arxiv.org/abs/0804.4527 (2008).
\bibitem{fu2008} L. Fu and C.L. Kane, Phys. Rev. Lett. {\bf 100}, 096407 (2008).
\bibitem{murakami06} S. Murakami, Phys.\ Rev. Lett. {\bf 97}, 236805 (2006).
\bibitem{disalvo99} F.J. DiSalvo, Science {\bf 285}, 703 (1999).
\bibitem{wyckoff86} R.W.G. Wyckoff, \emph{Crystal Structures} (Krieger, Malabar, 1986).
\bibitem{shroeder73} B. Schroeder \emph{et al.}, Phys.\ Stat. Sol. B {\bf 59}, 561 (1973).
\bibitem{black57} J. Black \emph{et al.}, J. Phys. Chem. Solids {\bf 2}, 240 (1957).
\bibitem{larson02} P. Larson \emph{et al.}, Phys.\ Rev. B {\bf 65}, 085108 (2002).
\bibitem{mishra96} S.K. Mishra \emph{et al.}, J. Phys.: Condens. Matter {\bf 9}, 461 (1997).
\bibitem{greanya02} V.A. Greanya \emph{et al.}, J. App. Phys. {\bf 92}, 6658 (2002).
\bibitem{hyde74} G.R. Hyde \emph{et al.}, J. Phys. Chem. Solids {\bf 35}, 1719 (1974).
\bibitem{kohler75} H. Kohler and A. Fabricius, Phys.\ Status Solidi B {\bf 71}, 487 (1975).
\bibitem{blaha01} P. Blaha  \emph{et al.}, computer code WIEN2K (Vienna University of Technology, Vienna, 2001).
\bibitem{perdew96} J.P. Perdew  \emph{et al.}, Phys.\ Rev. Lett. {\bf 77}, 1996 (1996).
\bibitem{heske99} C. Heske  \emph{et al.}, Phys.\ Rev. B {\bf 59}, 4680 (1999).
\bibitem{lashell96} S. LaShell  \emph{et al.}, Phys. Rev. Lett. {\bf 77}, 3419 (1996).
\bibitem{hoesch04} M. Hoesch  \emph{et al.}, Phys. Rev. B {\bf 69}, 241401 (2004).
\bibitem{ryu} A.P. Schnyder  \emph{et al.}, Phys. Rev. B {\bf 78}, 195125 (2008).
\bibitem{sbarpes} K. Sugawara \emph{et al.}, Phys. Rev. Lett. {\bf 96}, 046411 (2006).
\bibitem{hsieh09} D. Hsieh, M.Z. Hasan  \emph{et al.}, submitted.



\end{thebibliography}
\end{document}